\documentclass[manuscript,screen]{acmart}
\AtBeginDocument{%
  }

\setcopyright{acmlicensed}
\copyrightyear{2018}
\acmYear{2018}
\acmDOI{XXXXXXX.XXXXXXX}
\acmConference[Conference acronym 'XX]{Make sure to enter the correct
  conference title from your rights confirmation email}{June 03--05,
  2018}{Woodstock, NY}
\acmISBN{978-1-4503-XXXX-X/2018/06}

\usepackage[caption=false,font=normalsize,labelfont=sf,textfont=sf]{subfig}
\usepackage{textcomp}
\usepackage{stfloats}
\usepackage{url}
\usepackage{verbatim}
\usepackage{cite}
\usepackage{booktabs} 
\usepackage{amsmath,amsfonts}

\usepackage{pifont}
\usepackage{bm}
\usepackage{mathrsfs}

\usepackage{algpseudocode}
\usepackage[linesnumbered,ruled,vlined]{algorithm2e}
\SetKwInOut{KwIn}{Input}
\SetKwInOut{KwOut}{Output}

\usepackage{graphicx}
\graphicspath{{fig/}}

\usepackage{multirow}
\newcommand{\tabincell}[2]{\begin{tabular}{@{}#1@{}}#2\end{tabular}}
\usepackage{tabularx}
\usepackage{rotating}

\usepackage{color}

\usepackage{enumitem}

\usepackage{lineno,hyperref}

\begin{document}

\title{Multi-Agent Coordination across Diverse Applications: A Survey}

\author{Lijun Sun}
\email{sunlijun@sztu.edu.cn}
\affiliation{%
  \institution{Shenzhen Technology University}
  \country{China}
}

\author{Yijun Yang}
\email{altmanyang@tencent.com}
\affiliation{%
  \institution{Tencent}
  \country{China}
}

\author{Qiqi Duan}
\email{11749325@mail.sustech.edu.cn}
\author{Yuhui Shi}
\email{shiyh@sustech.edu.cn}
\affiliation{%
  \institution{Southern University of Science and Technology}
  \country{China}
}

\author{Chao Lyu}
\authornote{Corresponding author.}
\email{lyuchao@swu.edu.cn}
\affiliation{%
  \institution{Southwest University}
  \country{China}
}

\author{Yu-Cheng Chang}
\email{Fred.Chang@uts.edu.au}
\author{Chin-Teng Lin}
\email{Chin-Teng.Lin@uts.edu.au}
\author{Yang Shen}
\authornotemark[1]
\email{Yang.Shen-9@student.uts.edu.au}
\affiliation{%
 \institution{University of Technology Sydney}
 \country{Australia}
}

\renewcommand{\shortauthors}{Sun et al.}

\begin{abstract}
Multi-agent coordination studies the underlying mechanism enabling the trending spread of diverse multi-agent systems (MAS) and has received increasing attention, driven by the expansion of emerging applications and rapid AI advances.
This survey outlines the current state of coordination research across applications through a unified understanding that answers four fundamental coordination questions: (1) what is coordination; (2) why coordination; (3) who to coordinate with; and (4) how to coordinate.
Our purpose is to explore existing ideas and expertise in coordination and their connections across diverse applications, while identifying and highlighting emerging and promising research directions. 
First, general coordination problems that are essential to varied applications are identified and analyzed.
Second, a number of MAS applications are surveyed, ranging from widely studied domains, e.g., search and rescue, warehouse automation and logistics, and transportation systems, to emerging fields including humanoid and anthropomorphic robots, satellite systems, and large language models (LLMs).
Finally, open challenges about the scalability, heterogeneity, and learning mechanisms of MAS are analyzed and discussed.
In particular, we identify the hybridization of hierarchical and decentralized coordination, human-MAS coordination, and LLM-based MAS as promising future directions.
\end{abstract}



\received{2025}

\maketitle

\section{Introduction}

\begin{figure*}[htbp!]
\centering
\includegraphics[width=0.98\linewidth]{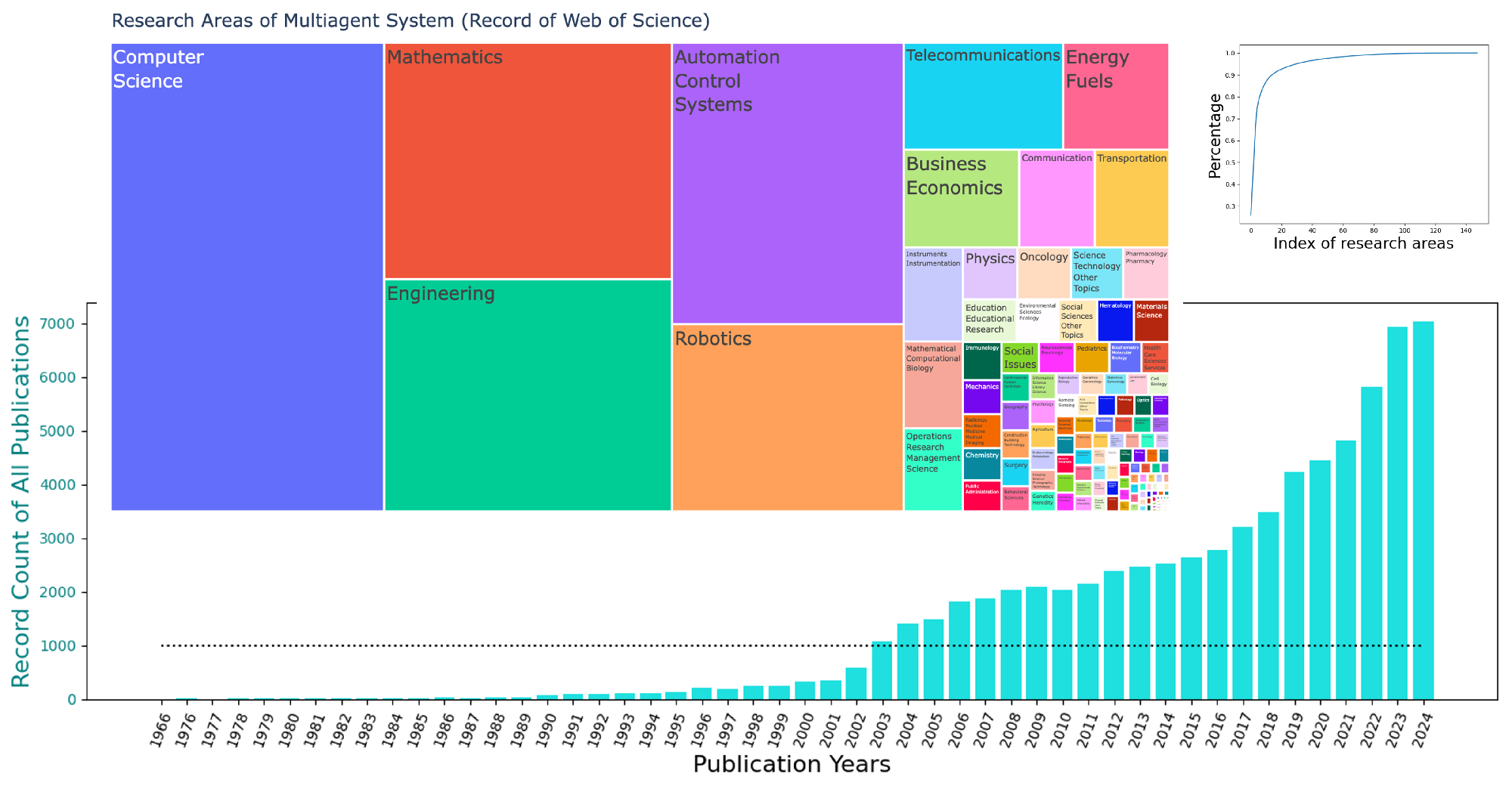}
\caption{The number of publications and research areas of multi-agent system (MAS) research based on the records of Web of Science (WOS).
The MAS topic covers 148 of a total of 252 research areas.
The record count determines the rectangular size for each research area, with the top 15 being: Computer Science, Mathematics, Engineering, Automation Control Systems, Robotics, Telecommunications, Energy Fuels, Business Economics, Communication, Transportation, Instruments Instrumentation, Mathematical Computational Biology, Operations Research Management Science, Physics, and Oncology.
}
\label{fig_mas_publication}
\end{figure*}
%


In the past 30 years, multi-agent systems (MAS) have gained growing interest from academics and industries as an interdisciplinary research topic, as shown in Fig.~\ref{fig_mas_publication}.
This spreading trend of MAS has especially accelerated in the last decade.
New MAS applications and multi-agent tasks emerge with technological advancement and new requirements, which offer novel opportunities and challenges to better solve complex problems from the MAS perspective.
Most recently, multiple large language model (LLM) based MAS methods have demonstrated the desirable ability of collective (a.k.a. swarm) intelligence in solving complex and challenging problems with human-like capabilities, such as reasoning and planning in natural languages.
Multi-agent coordinated autonomous driving (AD) seems to be a not-so-far future with more AD vehicles of increasing automation running on road~\citep{litman2024autonomous}.
Therefore, further study on the cross-domain MAS is necessary to promote knowledge inspiration and transfer across applications.

In this context, multi-agent coordination is the shared key mechanism that underlies the system integration and a force multiplier of MAS.
Wooldridge~\citep{wooldridge2009introduction} describes it as the defining problem in working together.
Particularly, coordination is defined as managing dependencies of agents' activities by Malone et al.~\citep{malone1994interdisciplinary}, which forms a cornerstone of the interdisciplinary coordination theory. 
On this base, the pervasive clustering phenomenon in MAS can be more easily explained as stemming from the spatio-temporal distribution of agents' multi-level dependencies, like the dynamic relationships in the first-order and higher-order attention mechanisms.
Moreover, clustering impacts coordination both logically and physically, including coordinated learning, communication, task allocation, consensus achievement, coalition/team/cluster/group formation, etc.
Therefore, the common clustering of agents is actually answering a very fundamental coordination question: \textit{who to coordinate with}.
Then, a subsequent fundamental coordination question is: \textit{how to coordinate}, i.e., managing dependencies.
It is the crucial part of coordinated decision-making and technically relates to learning, adaption, game theory, search, optimization, etc.

Out of the flourishing results, large amounts of MAS surveys classify coordination algorithms by techniques or tasks, or summarize the tasks of specific algorithmic techniques,  
such as the 
consensus algorithms~\citep{olfati2007consensus}, 
multi-agent planning approaches~\citep{torreno2017cooperative}, 
general multi-agent reinforcement learning (MARL) algorithms~\citep{zhang2021multi},
application domains of MARL~\citep{zhou2023multi},
MARL algorithms in Internet~\citep{li2022applications}, 
and autonomous driving (AV) tasks~\citep{mariani2021coordination}.
However, on one hand, rare survey works explicitly unify the interdisciplinary coordination research from the perspective of ``\textit{who to coordinate with}" and ``\textit{how to coordinate}", although they need to be generally addressed by every coordination process. 
On the other hand, commonalities of coordination mechanisms across diverse applications are always expected for summarizing the current and illuminating emerging/future theory and application directions.

\begin{figure*}[htbp!]
\centering
\includegraphics[width=0.88\linewidth]{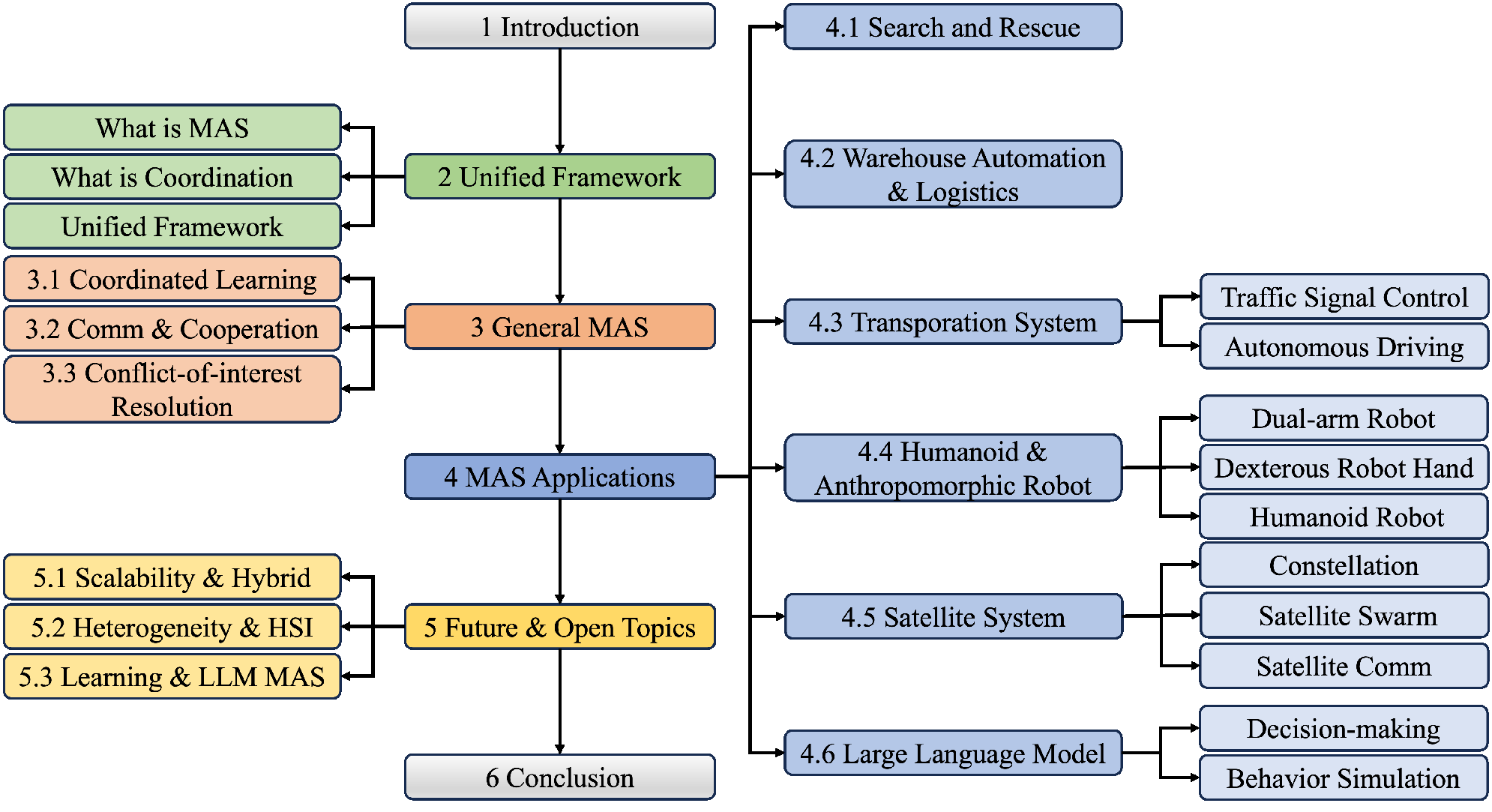}
\caption{The structure of this survey.
A unified framework is introduced in Section~\ref{sec_framework}.
Coordination problems for general MAS are reviewed in Section~\ref{sec_mas_general}.
MAS applications are surveyed in Section~\ref{sec_application}.
Future and open research topics are discussed in Section~\ref{sec_challenge}.
}
\label{fig_mas_outline}
\end{figure*}

Against this background, we explore a unified understanding of multi-agent coordination across applications.
To this end, we first explain ``\textit{what is coordination}" and ``\textit{why coordination}" in diverse contexts (i.e., coordinated tasks/problems in applications) throughout the survey.
Then, we describe the two fundamental questions: ``\textit{who to coordinate with}" and ``\textit{how to coordinate}" in one unified computational framework in Section II.
From this unified perspective, the commonalities and specialties of general coordination tasks and concrete coordination applications are captured in Section III and IV, respectively.
In particular, Section III answers ``\textit{how to coordinate}" from the methodology perspective for three general multi-agent tasks, while Section IV answers ``\textit{how to coordinate}" from the task perspective for six MAS applications.
Furthermore, challenges and promising directions are identified and discussed in Section V.
Finally, the conclusions are given in Section VI.
The outline of this survey is shown in Fig.~\ref{fig_mas_outline}.


%

\section{A Framework for Multi-agent Coordination}
\label{sec_framework}

\begin{figure*}[htbp!]
\centering
\includegraphics[width=0.9\linewidth]{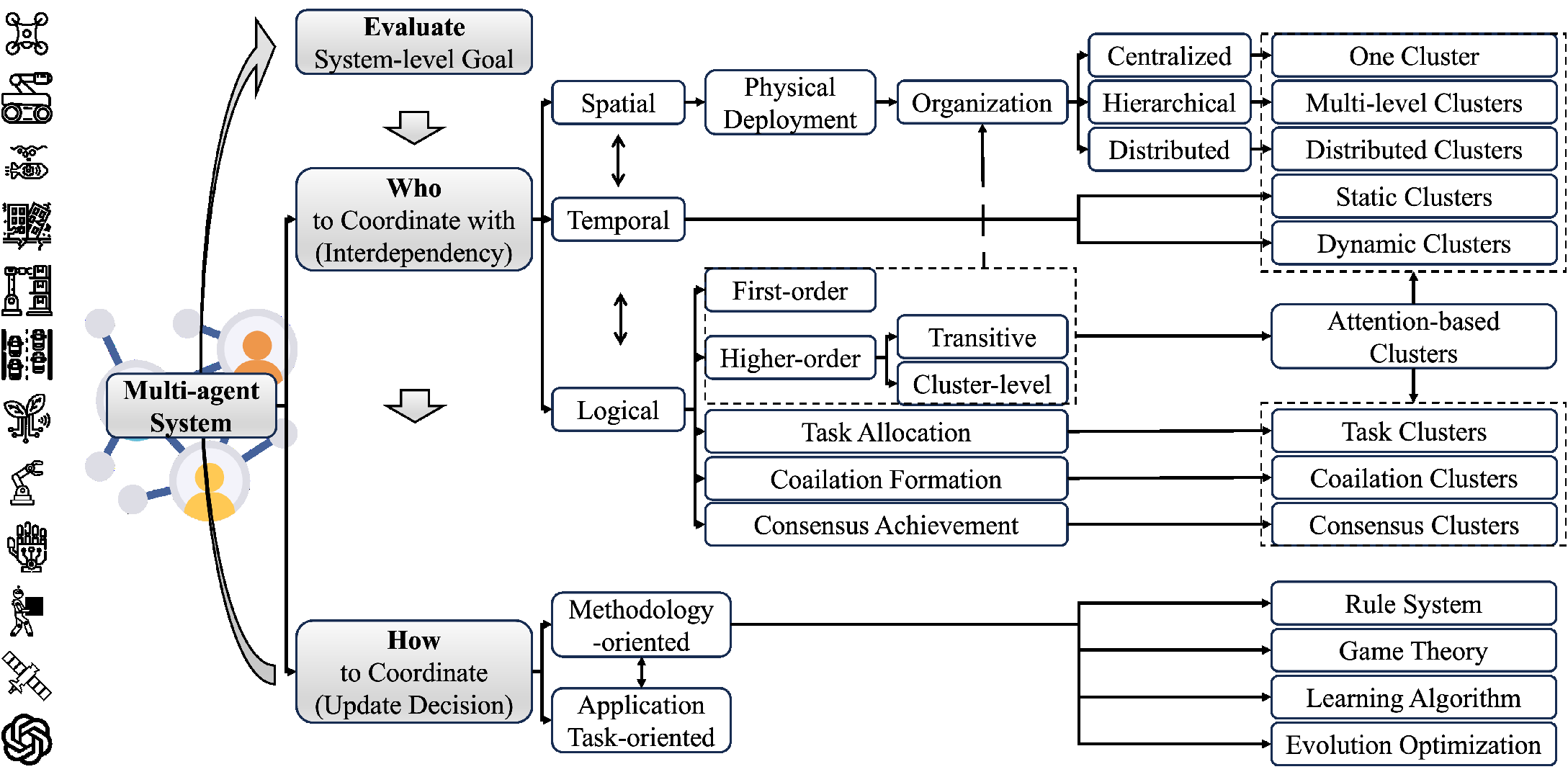}
\caption{The unified framework (perspective) of coordination in this survey. 
The coordination in sequential decision-making is an iterative process consisting of three components: evaluate the system-level performance, social choice on who to coordinate with, and how to coordinate. (Section~\ref{sec_framework})}
\label{fig_mas_unified}
\end{figure*}

\subsection{Multi-Agent Systems}
The definition of multi-agent systems can be as simple as a collection of autonomous agents with common or conflicting interests and information~\citep{shoham2008multiagent}.
The community of distributed artificial intelligence (DAI) concerns the concurrent artificial intelligence (AI) computations and inherent coordination problems of multiple agents~\citep{bond1988readings}.
Russell et al.~\citep{Russell2021Artificial} describe the multi-agent system from the agent's perspective that an agent's performance depends on other entities, which can be treated as other agents apart from the environment.
Here, the following definition is sufficient for our purpose, which is consistent with the one in~\citep{wooldridge2009introduction}.

\textit{\textbf{Definition 1.} The \textbf{multi-agent system} is a system consisting of multiple independent interactive decision makers called agents, where an agent may be a person, a robot, a robotic subsystem, a manipulator's finger, a distributed computing unit, a language model, a satellite, etc.}

\subsection{Coordination}
Coordination is an interdisciplinary concept with diverse definitions.
For example, Malone et al.~\citep{malone1994interdisciplinary} gives the definition that ``coordination is managing dependencies between activities" (tasks) of ``actors" (agents).
Generally, the terms ``coordination", ``cooperation" and ``collaboration" are conceptually different though similar, where the latter two can be seen as approaches to achieve or different forms of the first one.
Typical fundamental components in a coordination process are the agents (to be coordinated), overall system-level performance objectives (metrics), and often conflicts of (individual) interests.
Therefore, we give the following definition of the multi-agent coordination.

\textit{\textbf{Definition 2.} We present the \textbf{multi-agent coordination} as agents interact and make decisions for the overall system-level performance, including resolving their conflicted interests.
In particular, agents make two essential decisions: who to coordinate with and how to coordinate.}

\textbf{Unified framework.}
The whole coordination process in sequential decision making can be unified in an iterative process consisting of three components: evaluate system-level performance, social choice on who to coordinate with, and how to coordinate.
This unified framework is presented in Fig.~\ref{fig_mas_unified}.

\textbf{Who to coordinate with.}
In the above definition, the first decision problem ``who to coordinate with" determines the clusters of agents in terms of their inter-dependencies, such as the meta-level structure ~\citep{durfee1988coordination}, the coalition~\citep{jennings1995controlling}, and the structured groups of agents~\citep{gasser1988implementing}. 
These dependencies may be spatio-temporal different and of different orders, which can be the physical inter-agent spatial distances or the logical decisions.
For example, the organization structure of a MAS is a coordination graph that depicts the inter-agent dependencies.
In a centralized MAS, all agents communicate with the control center while the center decides which agent coordinates with which other agents, e.g., to form neighborhoods or resolve conflicts.
In this case, there is only one cluster in the MAS and the control center is the cluster center.
In a hierarchical or decentralized MAS, partial or all agents need to reason by themselves their relevant dependencies, where multi-level or distributed clusters emerge.
Take another example.
In the task allocation, inter-dependent agents are clustered based on the allocated task.
For instance, in the multi-target pursuit, a target is a task and a cluster center, while pursuing the same target triggers the intensive intra-cluster coordination of (pursuer) agents.

Besides the first-order inter-agent dependencies, higher-order dependencies also play a crucial role in answering ``who to coordinate with".
Take the transitive inter-agent dependency in forming the local coordination clusters as an example.
Suppose agent A is influenced by agent B, while agent B is influenced by agent C. 
Then, the resolution of coordination issue between A and B needs to involve C. 
A local coordination cluster is formed among A, B, and C through their transitive inter-agent dependencies.
A further example is modeling the cluster-level dependencies of agents to build a more refined coordination graph topology, such as more sparse yet more accurately weighted coordination relationships.

\textbf{How to coordinate.}
After characterizing the inter-dependencies and clustering agents, the second decision problem is ``how to coordinate".
It corresponds to ``manage dependency" in Malone's coordination definition and is the central part of coordinated decision-making algorithms~\citep{kochenderfer2022algorithms}.
Its mechanisms are often inspired by the survival and high efficiency of biological multi-agent systems, such as the self-organizing~\citep{camazine2001self} and social learning~\citep{fisher1949opening}.
From the perspective of methodologies, it is usually categorized by the rule-based methods (e.g., lexicographic order), game theory, learning-based approaches (e.g., multi-agent reinforcement learning), evolution-based schemes (e.g., multi-objective optimization), etc. 
From the perspective of application tasks, its taxonomy is typically domain-specific, which, however, relates with the combination of the above classifications, such as the centralized solver for resource sharing of heterogeneous swarm based on MARL.

\textbf{Evaluate system-level performance.}
The coordination performance is evaluated at a system level, for example, a balance or trade-off between individual agents' interests.
The global emerged intelligence is usually explained by the collective intelligence~\citep{malone2015handbook}, swarm intelligence~\citep{dorigo1999swarm,bayindir2016review}, or society of mind~\citep{minsky1988society}.

Finally, our unified understanding of the coordination across applications is inspired by the survival and high efficiency of biological multi-agent systems, including the human society ~\citep{camazine2001self}.
The emerged intelligence is usually explained by the society of mind~\citep{minsky1988society}, collective intelligence~\citep{malone2015handbook}, or swarm intelligence~\citep{dorigo1999swarm,bayindir2016review}.
In particular, the unified framework in Fig.~\ref{fig_mas_unified} resembles the brainstorming process of human coordination, where groups are formed by clustering and agreement on better solutions are asymptotically reached through repeated and intensive intra-cluster and inter-cluster interactions driven by the overall system-level performance.
It has inspired the brain storm optimization (BSO)~\citep{shi2015optimization} and mindstorm of large language model (LLM)~\citep{zhuge2023mindstorms} with inspiring results, which provide substantial evidence for its powerful computational intelligence.


\section{General MAS}
\label{sec_mas_general}
\begin{figure}[ht]
\centering
\includegraphics[width=0.65\linewidth]{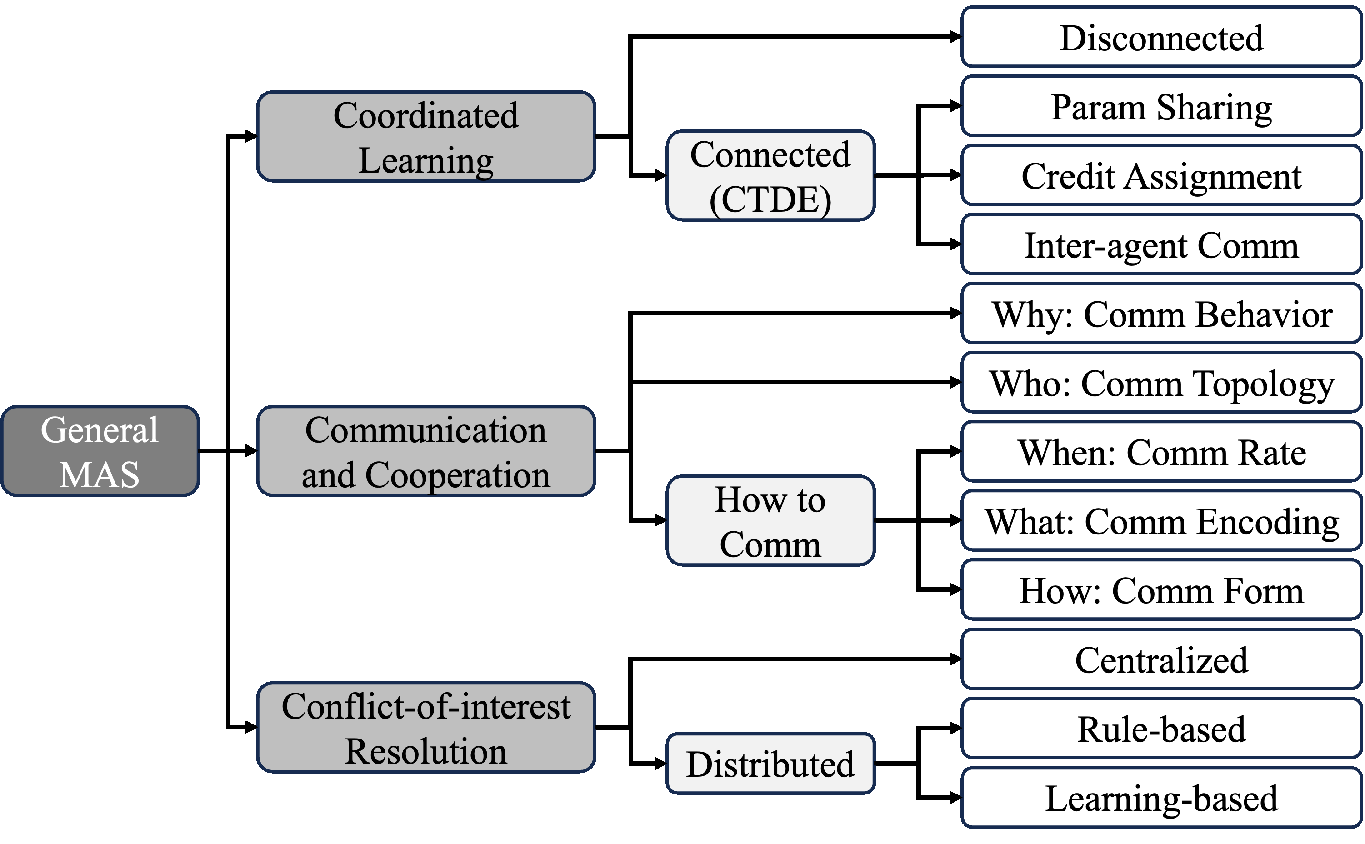}
\caption{General multi-agent tasks. (Section~\ref{sec_mas_general})}
\label{fig_mas_general}
\end{figure}

Generally, a multi-agent system forms a coordination graph where an agent is a node, and an edge represents some kind of interaction or relationship, such as learning, observing, communicating, cooperating, competing, collaborating, or other dependencies.
Based on the coordination topology, clusters emerge from some common interests of highly related and dependent agents, such as the same target or mutual benefits. As highlighted in the framework proposed in Section~\ref{sec_framework}, evaluation and clustering determine which agents coordinate with which agents, while updating determines how to coordinate. This section reviews several coordination tasks that are general and important to almost all applications of MAS, which is outlined in Figure~\ref{fig_mas_general} and summarized in Table~\ref{tab:mas_general}.

In a MAS, coordination refers to the process by which multiple agents work together, communicate, and adjust their actions to achieve a common goal. Coordination ensures that agents can effectively cooperate, avoid conflicts, and optimize the overall performance of the system by harmonizing their behaviors. The coordinated learning among agents is discussed in Section~\ref{sec_coordinated_learning}, communicating and cooperating is discussed in Section~\ref{mas_coop}, and avoiding conflicts is discussed in Section~\ref{mas_conflict}.

\begin{table*}[htbp!]
\centering
\footnotesize
\setlength{\tabcolsep}{3pt}
\caption{General Multi-Agent System (MAS) from the unified perspective. (Section~\ref{sec_mas_general})}
\label{tab:mas_general}
\begin{tabular}{lllll}
\toprule
\textbf{General MAS}
& \multicolumn{2}{l}{\textbf{Unified perspective}}
& \multicolumn{2}{l}{\textbf{Paper}} 
\\
\midrule
\multirow{8}{*}{\textbf{\tabincell{l}{Coordinated \\Learning (CL)}}}
& \multicolumn{3}{l}{Why CL (Benefits)} &~\citep{lowe2017multi},~\citep{tan1993multi,yu2022surprising}  
\\ \cline{2-5}
{} 
& \multirow{4}{*}{Who to Learn from} 
& \multirow{3}{*}{Inter-agent Comm} 
& Disconnected 
& Independent RL 
\\
{} & {} & {} 
& Fully Connected 
& MAPPO~\citep{yu2022surprising}, VDN~\citep{sunehag2018value}, QMIX~\citep{rashid2020monotonic} 
\\
{} & {} & {} 
& Sparsely Connected 
& CommNet~\citep{sukhbaatar2016learning} 
\\
{} & {} & {} 
& Higher-order Relation
& LTS-CG~\citep{duan2024inferring}, GACG~\citep{duan2024group} 
\\ \cline{2-5}
{} & \multirow{3}{*}{How to Learn} 
& \multicolumn{2}{l}{Parameter Sharing} 
& ~\citep{Gupta2017}  
\\
{} & {} 
& \multicolumn{2}{l}{Credit Assignment} 
& MADDPG~\citep{lowe2017multi}, COMA~\citep{foerster2018counterfactual}, IC3Net~\citep{singhlearning} 
\\
{} & {} 
& Inter-agent Comm 
& Learned Content 
& DIAL~\citep{foerster2016learning}, BiCNet~\citep{peng2017multiagent}, SchedNet~\citep{kimlearning} 
\\ \midrule
\multirow{8}{*}{\textbf{\tabincell{l}{Communication \\\& Cooperation}}}
& \multicolumn{2}{l}{Why Comm (Motivation)} 
& \multicolumn{2}{l}{~\citep{Russell2021Artificial}, Convergence~\citep{olfati2007consensus, nedic2018network}, Event-trigger~\citep{nowzari2019event, ge2020dynamic}} 
\\ \cline{2-5}
{} 
& Who to Comm with 
& Who: Comm Topology 
& \multicolumn{2}{l}{\tabincell{l}{Deployment, Heuristic, Optimization~\citep{wu2008construction, meng2019deep}, Adaptive~\citep{jiang2018learning, du2021learning}, \\Broadcast+Attention~\citep{hoshen2017vain, das2019tarmac}}}
\\ \cline{2-5}
{} 
& \multirow{4}{*}{How to Comm} 
& When: Comm Rate
& \multicolumn{2}{l}{Learning-based~\citep{singhlearning,kimlearning}} 
\\ 
{} & {} 
& What: Comm Encoding
& \multicolumn{2}{l}{Stigmergy~\citep{theraulaz1999brief}, Evolution-based~\citep{giles2003learning}, Learning-based~\citep{sukhbaatar2016learning, foerster2016learning}}
\\ 
{} & {} 
& How: Comm Form 
& \multicolumn{2}{l}{\tabincell{l}{Broadcast, Multi-hop, Point-to-point, Synchronous, Asynchronous, \\Hierarchical}}
\\ \midrule
\multirow{6}{*}{\textbf{\tabincell{l}{Conflict-of-interest\\Resolution}}}
& \multicolumn{2}{l}{Why Coordinate (Analysis)} 
& \multicolumn{2}{l}{Shared resource, Optimize overall goal,~\citep{gao2023review}} 
\\ \cline{2-5}
{} 
& Who to Coordinate with 
& Centralized Solver
& \multicolumn{2}{l}{\tabincell{l}{Meta-agent~\citep{sharon2012meta}, Conflict graph~\citep{felner2018adding}, Dependency graph~\citep{li2019improved}, \\Others~\citep{boyarski2015icbs}}}
\\ \cline{2-5}
{} 
& \multirow{3}{*}{How to Coordinate} 
& Priority/Rule-based
& \multicolumn{2}{l}{Lexicographic Convention~\citep{boutilier1996planning, sun2023toward}}
\\ 
{} & {} 
& Distributed Learning-based
& \multicolumn{2}{l}{\tabincell{l}{Prioritized comm~\citep{ye2022multi, li2022multi}, Scalable comm~\citep{wang2023scrimp}, \\Higher-order relation (attention)~\citep{guan2022ab}, IL+RL+Rule~\citep{sartoretti2019primal, li2020graph}}}
\\
\bottomrule
\end{tabular}
\end{table*}

\subsection{Coordinated Learning}
\label{sec_coordinated_learning}
The coordinated learning (CL) or social learning of agents answer two questions from the unified perspective: who to learn from and what to learn.
In the CL graph, an agent is a node, and an edge exists if there are learning interactions by exchanging information.
Extending single-agent reinforcement learning (RL) directly to multi-agent settings makes agents learn independently and is, therefore, called independent learning.
It is a disconnected, coordinated learning graph.
Though simple, its main issue is the non-stationary problem~\citep{lowe2017multi}.
In contrast, multi-agent coordinated learning is achieved by utilizing some kind of global state, which may be the ground truth provided by the environment or based on explicit communications that serve as a learning purpose in training (yet such comms are not restricted only in training, but can also be in execution).
Past works have proved its benefits in learning efficiency~\citep{tan1993multi,yu2022surprising}.

Centralized training and decentralized execution (CTDE) is a typical coordinated learning paradigm.
In the CTDE, agents (policies) work as parallel and decentralized executives collecting sufficient experiences to be communicated to a training center, while the training center works as a data center and learning center where the critic function may access additional information, in which case we call it a centralized critic function.
Moreover, the CTDE is often combined with the parameter sharing~\citep{Gupta2017}, where a shared (centralized) critic function and, optionally, a shared policy function are centralized learned.
Therefore, in the centralized learning of CTDE, agents can be fully connected in terms of coordinated learning, such as multi-agent actor-critic methods MAPPO~\citep{yu2022surprising}, and value-decomposition methods VDN~\citep{sunehag2018value} and QMIX~\citep{rashid2020monotonic}, although some of which also support partial connections.
Particularly, the credit assignment problem can be alleviated in several ways for heterogeneous agents, such as the works of MADDPG~\citep{lowe2017multi}, COMA~\citep{foerster2018counterfactual}, and IC3Net~\citep{singhlearning}.
In addition to the predefined communication contents like local observations, actions, or shared parameters between agents and the training center in CTDE, the coordinated learning can be augmented with direct inter-agent communication protocols that can be learned, such as DIAL~\citep{foerster2016learning}, BiCNet~\citep{peng2017multiagent}, and SchedNet~\citep{kimlearning}, which will be kept after training.
Although some of these works use the broadcasting communication channel, sparse communication may be learned, such as CommNet~\citep{sukhbaatar2016learning}.
Besides, an emerging technique is to model higher-order interaction relationships, such as group-level (or cluster-level) dependencies, along with agent-level interactions in the coordination graph for sparse and weighted inter-agent communications, e.g. LTS-CG~\citep{duan2024inferring}, GACG~\citep{duan2024group}.
More related work in terms of communication will be introduced in the next topic.

\subsection{Communication and Cooperation}
\label{mas_coop}

Communication and cooperation are the backbone of MAS, enabling agents to work together effectively toward shared goals. 
Without communication and cooperation, agents cannot share vital information or synchronize their actions, leading to inefficiencies, conflicts, and resource mismanagement. 
Therefore,  communication can be seen as a rational behavior for coordination~\citep{Russell2021Artificial}.
Coordination ensures that agents operate in harmony, optimizing task allocation, resolving conflicts, and enhancing the system's adaptability and robustness.

The communication network topology can be seen as a graph $G=(V,E)$, where each agent is a node $v_i\in V$, and an edge $e_{ij}\in E$ means there are communications between agent $i$ and agent $j$. The connections among agents form clusters for coordination. However, communication resources are often constrained, which influences the feasibility of coordination algorithms and partly motivates the selective communication as one optimization objective. This part's literature can be categorized by answering two key questions of coordination: which agents to communicate and cooperate with and how agents communicate and cooperate with each other.

{\bf Who.} The communication topology is often a key factor for designing coordination algorithms and identifying conditions for convergence (rate) of multi-agent optimization~\citep{olfati2007consensus,nedic2018network}. The selective topology can be the result of physical deployment, heuristic design, topology optimization~\citep{wu2008construction, meng2019deep}, or adaptive topology control~\citep{jiang2018learning,du2021learning}. 
Besides, this selective communication can also be achieved by weighting the incoming messages without altering the binary edge connection, such as using attention weights~\citep{bahdanau2015neural} along with broadcasting~\citep{hoshen2017vain,das2019tarmac}. However, such attentional implementation will not reduce the communication overhead. 
From the unified perspective, a key point of the coordination in selective communication is that evaluation and clustering behavior help agents determine which agents to communicate and cooperate with.

{\bf How.} The other key point of communication and coordination is how agents communicate among themselves, which can be concluded in the following aspects. First, in terms of the communication rate, the two most commonly used schemes are time-triggered and event-triggered communication. Event-triggered communication is proposed to trigger the data transmission by event(s). Compared with time-triggered or periodic schemes, it can improve efficiency, flexibility, and scalability~\citep{nowzari2019event} with the cost of a potentially compromised coordination performance~\citep{ge2020dynamic}. 
Besides explicit predefined event triggers, when to trigger communication can also be learned~\citep{singhlearning,kimlearning}. 
Second, effective communication and coordination often depend on the message content, which could be either direct (via message passing) or indirect (via stigmergy~\citep{theraulaz1999brief}, i.e., by changing the environment). 
On the other hand, regarding message encoding, the communication message can be encoded by evolution~\citep{giles2003learning} or learning~\citep{sukhbaatar2016learning,foerster2016learning} apart from explicitly predefined. 
Last but not least, communication and coordination among agents can be in different forms, including broadcasting, multi-hop, point-to-point, synchronous, asynchronous, hierarchical, etc. In a MAS, communication and cooperation cause updates to the system, which corresponds to ``\textit{how to coordinate}" in the proposed unified framework of Fig.~\ref{fig_mas_unified}.





\subsection{Conflict-of-interest Resolution}
\label{mas_conflict}


Conflict resolution in multi-agent coordination refers to the process of managing and resolving situations where multiple agents attempt to access shared resources or perform actions in a way that leads to conflicts or collisions, e.g., occupy the same space. Efficient conflict resolution ensures smooth, coordinated behavior while meeting the system’s overall goals. An individual agent's own goal represents its interest, while conflicts of interest may occur during interactions among agents and need to be coordinated. Popular resolution mechanisms include path planning (agents plan their paths to avoid conflicts or collisions), resource allocation (agents take turns to access shared resources based on scheduling or priority), behavioral adjustments (agents modify their actions, such as stopping, waiting, or re-routing), etc.

Coordination issues may cause clustering behaviors of physical collisions, deadlocks \footnote{Deadlock: agents stop moving forever as if being locked before final goals are reached.}, and live-locks \footnote{Live-lock: agents can move but are coupled (locked) with each other and fail to make further progress.}.
In particular, collisions may be catastrophic in safety-critical scenarios such as automated warehouses and transportation systems where agents have physical forms (e.g., robots or vehicles). 
Collision resolution is generally based on some kind of priority, i.e., which conflict to resolve first and which agent should adjust its behavior or yield. 
As an example of priority-based conflict resolution, it deserves to mention the lexicographic convention~\citep{boutilier1996planning} in the distributed immediate coordination due to its simplicity, generality, and effectiveness in guaranteeing the safety~\citep{sun2023toward}.
However, a common problem of rule-based solutions is that they are hard to provide optimal solutions for all cases in spite that they are often designed case-by-case.
Take the Multi-Agent Path Finding (MAPF) problem as an example.
The priority of (path) conflicts significantly matters in terms of the efficiency and effectiveness performance~\citep{gao2023review}, which may take the form of meta-agent (group of heavily conflicting agents)~\citep{sharon2012meta}, conflict graph~\citep{felner2018adding}, dependency graph~\citep{li2019improved}, and other schemes~\citep{boyarski2015icbs} for centralized MAPF solvers. 
For distributed learning-based MAPF solvers, collision with agents and even obstacles is still a big problem, although different communication schemes are explicitly used, such as prioritized communication~\citep{ye2022multi, li2022multi}, scalable communication~\citep{wang2023scrimp}, and higher-order relationships by attention mechanism~\citep{guan2022ab}. 
Besides, the combination of imitation learning (IL) and reinforcement learning-based policy with rule-based collision avoidance mechanism may lead to deadlocks since the protective compulsory behavior is out of the consideration of learned policy and thus results in uncoordinated cases~\citep{sartoretti2019primal, li2020graph}.

As a sum-up, the guaranteed safety in resolving the conflicts of interest in MAS faces the curse of dimensionality for centralized solvers, the imperfection of rule-based distributed solutions, and the immaturity of distributed learning-based methods.

\section{MAS Applications}
\label{sec_application}

This section presents a non-exhaustive list of MAS applications aiming to illustrate its wide spectrum in a unified view, as shown in Table~\ref{tab:mas_application}.
We choose the widely-studied applications largely inspired by the topic areas of MAS surveys in the last five years, which illuminate the contemporary researchers' attention, and highlight emerging and promising applications.

Multi-agent systems, multi-robot systems (MRS), or swarm systems have proposed new autonomous and intelligent solutions to traditional and beyond traditional applications, such as the
multi-robot multi-station system in manufacture~\citep{zhou2022multi},
robotic sorters in recyclable industrial waste~\citep{kiyokawa2022challenges},
warehouse robots in logistics~\citep{da2021robotic},
service and assistance robots in health care (see~\citep{benavidez2015design,papadakis2018system,barber2022multirobot} and references therein),
nanorobotics in precision medicine~\citep{rajendran2024nanorobotics},
and swarm robots in many future applications~\citep{dorigo2021swarm}. 
In more detail, an agent or a robot can be an 
unmanned aerial/ground/underwater vehicle (UAV/UGV/UUV)~\citep{ahmed2022recent,jamshidpey2024centralization,zhou2021survey},
humanoid robot~\citep{saeedvand2019comprehensive},
assistive robot~\citep{broekens2009assistive} (e.g., smart wheelchair, exoskeleton, pet(-like) robot~\citep{bharatharaj2015bio}),
snake robot~\citep{liu2021review},
(rigid or soft) crawling/climbing robot~\citep{pshenin2022robot,chen2020soft,fang2022design}, 
sensor node~\citep{chung2011toward},
large language model (LLM)~\citep{zhuge2023mindstorms}, etc.
The following highlights several state-of-the-art coordination results that are implemented ready for or already used in real-world applications.

\begin{table*}[htbp!]
\centering
\footnotesize
\setlength{\tabcolsep}{3pt}
\caption{MAS applications from the unified perspective. (Section~\ref{sec_application})}
\label{tab:mas_application}
\begin{tabular}{llll}
\toprule
\multicolumn{2}{l}{\textbf{MAS Application}}
& \textbf{Unified perspective}
& \textbf{Paper} 
\\ \midrule
\multirow{8}{*}{\textbf{\tabincell{l}{Transportation\\Systems}}}
& {} 
& Why MAS \& Coordinate 
& \tabincell{l}{System-level goal (incl. shared resource)~\citep{balaji2010multi,  wu2021flow}, \\ Multi-agent nature (incl. problem decomposition)~\citep{balaji2010multi}}
\\ \cline{2-4}
{} 
& \multirow{2}{*}{Traffic Signal Control} 
& Who to Coordinate with 
& Intersection agent 
\\ 
{} & {} 
& How to Coordinate 
& Traffic signal timing 
\\ \cline{2-4}
{}
& \multirow{4}{*}{Autonomous Driving} 
& Who to Coordinate with 
& Intersection-Vehicle~\citep{hausknecht2011autonomous}, 
Fully connected AV~\citep{mariani2021coordination},
Spatial attention~\citep{liu2024attention} 
\\ \cline{4-4}
{} & {} 
& How to Coordinate 
& \tabincell{l}{Vehicle platooning~\citep{li2015overview,liang2015heavy}, 
Cruise control~\citep{desjardins2011cooperative}, 
Merging~\citep{shalev2016safe}, \\Navigating through traffic intersections~\citep{kalantari2016distributed}, Adversarial traffic~\citep{wachi2019failure}, \\Following, Lane changing, Overtaking~\citep{yu2019distributed} }
\\ \midrule
\multirow{10}{*}{\textbf{\tabincell{l}{Humanoid \&\\Anthropomorphic\\Robot}}} 
& {} 
& Why MAS \& Coordinate 
& \tabincell{l}{Incapacity of single agent~\citep{ollero2021past}, Dexterity ~\citep{ma2011dexterity}, Emulation~\citep{bullock2011classifying,feix2015grasp}, \\Scalability \& Adaptability~\citep{ren2024enabling, tao2023multi, ha2021learning}, Higher-order relationships} 
\\ \cline{2-4}
{} 
& \multirow{4}{*}{Dual-arm Robot} 
& Who to Coordinate with
& Grasp pair, Multiple arms 
\\ \cline{4-4}
{} & {} 
& How to Coordinate 
& \tabincell{l}{Combinatorial optimization, Multi-objective optimization~\citep{ren2024enabling}, \\Motion synchronization~\citep{asfour2006coordinated}, Inter-manipulator collision avoidance~\citep{das2020learning}, \\Decentralized control~\citep{ha2021learning} }
\\ \cline{2-4}
{} 
& \multirow{2}{*}{Dexterous Robot Hand} 
& Who to Coordinate with 
& Multi-finger 
\\
{} & {} 
& How to Coordinate 
& Centralized~\citep{andrychowicz2020learning}, Decentralized~\citep{wen2022multi,tao2023multi} 
\\ \cline{2-4}
{} 
& \multirow{2}{*}{Humanoid Robot} 
& Who to Coordinate with 
& Subsystems (incl. head, eye/vision, acoustics, multi-modal sensors) 
\\
{} & {} 
& How to Coordinate 
& Task decomposition \& transition, 
~\citep{kuang2012active},
~\citep{asfour2006coordinated}, 
Sensor fusion~\citep{mittendorfer2011humanoid}
\\ \midrule
\multirow{15}{*}{\textbf{\tabincell{l}{Satellite\\Systems}}}
& {} 
& Why Coordinate 
& 
\tabincell{l}{Mult-agent nature (distributed space system (DSS)), \\System-level performance optimization (incl. resource), ~\citep{radhakrishnan2016survey, kodheli2020satellite, farrag2021satellite, engelen2011systems},\\Scalable~\citep{zhang2021aser,lyu2024dynamic}, Upgradable, Robust, Low complexity, \\Service coverage \& continuity \& cost~\citep{savitri2017satellite, singh2020low, al2021optimal},\\
Mass production \& deployment, Reconfigurable} 
\\ \cline{2-4}
{} 
& \multirow{2}{*}{Constellation} 
& Who to Coordinate with 
& Satellites (in different orbital planes) 
\\ 
{} & {}
& How to coordinate 
& Many-objective optimization, Constrained optimization
\\ \cline{2-4}
{}
& \multirow{5}{*}{Satellite Swarm} 
& Who to Coordinate with 
& \tabincell{l}{Small satellites, Magnetic Nano-Probe Swarm~\citep{lubberstedt2005magnas}, QB50~\citep{QB50}, \\OLFAR~\citep{engelen2010olfar}} 
\\ \cline{4-4}
{} & {}
& How to Coordinate 
& \tabincell{l}{Self-organization mechanisms~\citep{sundaramoorthy2010systematic}, Mission scheduling \& planning~\citep{zheng2017swarm}, \\Consensus~\citep{sarlette2007cooperative}, Synchronization~\citep{marrero2022architectures}, Remote sensing~\citep{farrag2021satellite}, \\Collision avoidance~\citep{nag2013behaviour}, Target tracking \& navigation~\citep{stacey2018autonomous}} 
\\ \cline{2-4}
{}
& Satellite Comm 
& Who to Coordinate with 
& \tabincell{l}{Multi-beam satellite system~\citep{lin2022dynamic}, Phased array antenna~\citep{quadrelli2019distributed}, \\Multi-satellite multi-beam system~\citep{zhu2021satellite,lin2024satellite}}, 
\\ \cline{4-4}
{} & {} 
& How to Coordinate 
& Multi-agent beam hopping (scheduling)~\citep{lin2022dynamic},
Distributed routing~\citep{zhang2021aser,lyu2024dynamic} 
\\ \midrule
\multirow{12}{*}{\textbf{\tabincell{l}{LLM-based\\Multi-agent\\Systems}}} 
& {} 
& Why Coordinate 
& \tabincell{l}{Collective intelligence from diverse expertise, \\
Mimic human/animal group work \& behaviors,
Complex interactions}
\\ \cline{2-4}
{} 
& \multirow{3}{*}{Decision-making} 
& Who to Coordinate with 
&  \tabincell{l}{Role-playing~\citep{li2023camel}(CAMEL), ~\citep{qian2023communicative}, 
High-level communication~\citep{mandi2024roco},
\\Consensus~\citep{chen2023multi}}
\\ \cline{4-4}
{} & {} 
& How to Coordinate 
& \tabincell{l}{Collaborative programming~\citep{li2023camel},
Scientific research~\citep{zheng2023chatgpt},(CAMEL)~\citep{qian2023communicative},\\
Embodied intelligence~\citep{mandi2024roco}(RoCo: motion planning), \\ReAd: principled credit assignment~\citep{zhang2024towards}}
\\ \cline{2-4}
{}
& \multirow{4}{*}{Behavior Simulation} 
& Who to Coordinate with 
& \tabincell{l}{Social interaction: Human-agent interaction by NL~\citep{park2023generative}, \\Role-playing~\citep{ge2024scaling}, Social networks~\citep{gao2023s}}
\\ \cline{4-4}
{} & {} 
& How to Coordinate 
& \tabincell{l}{Game-playing~\citep{akata2023play}(Behavior game theory), 
Rationality analysis~\citep{DBLP:conf/aaai/FanCJ024},\\ Benchmarking: Welfare Diplomacy~\citep{mukobi2023welfare}, Multi-agent text game~\citep{li2023theory},\\Scale synthetic data generation~\citep{ge2024scaling}, Recommendation systems~\citep{agent4rec}}
\\
\bottomrule
\end{tabular}
\end{table*}

\subsection{Search and Rescue (SAR)}

Search and rescue (SAR) involves locating and assisting individuals in distress or facing imminent danger. The field encompasses several specialized areas, often defined by the terrain of the operation. Key types include mountain rescue for rugged areas, ground search and rescue, urban search and rescue for incidents in cities, combat search and rescue in battlefield settings, and air-sea rescue for operations over water. Multi-robot systems are useful tools for searching different types of environments. A recent survey has evaluated the current status of multi-robot systems in the context of search and rescue, for more details, please refer to~\citep{drew2021multi}.

There are many real-world multi-agent SAR applications. For example, swarm robots can be deployed to search for survivors in areas that are dangerous or inaccessible to humans, like earthquake zones or collapsed buildings. Besides, robots can track forest fires, floods, or other environmental hazards~\citep{bakhshipour2017swarm, arnold2018search}. In such tasks, the robots are expected to coordinate and sweep through the environment to search for the targets. A common way is to let the robots form a certain shape (e.g., line shape or V-shape).
Such shape formation behavior and sweeping through the environment correspond to ``\textit{who to coordinate with}" and ``\textit{how to coordinate}" in the unified coordination framework of Fig.~\ref{fig_mas_unified}.

\subsection{Warehouse Automation and Logistics}

MAS is widely used in warehouse automation and logistics, such as Amazon Robotics\footnote{\url{https://amazon.jobs/content/en/teams/ftr/amazon-robotics}} (formerly Kiva Systems), Cainiao smart warehouse\footnote{\url{https://www.cainiao.com/}}, drone delivery~\citep{dorling2016vehicle, frachtenberg2019practical}, etc. Autonomous agents, such as AGVs, manipulators, conveyors, and shuttles, coordinate to streamline tasks like picking, sorting, and transporting goods within the warehouse. Multi-agent coordination is essential for optimizing workflows and ensuring efficient material handling. Effective coordination allows these agents to communicate their positions, share task updates, negotiate resource usage, prevent bottlenecks, and minimize idle time. By working together seamlessly, agents can adapt to real-time changes in demand or inventory, dynamically reassigning tasks, and adjusting routes to maximize throughput. This level of coordination increases operational efficiency and scalability, making it easier to expand the system to handle larger volumes or more complex workflows. Typical coordination behaviors in such combinatorial challenges include task scheduling and path planning, as the competition tracks in the League of Robot Runners \footnote{\url{https://www.leagueofrobotrunners.org/}}.

\subsection{Transportation Systems}
\label{sec_transportation}
%

In the transportation system, two typical entities that can be modeled as agents are the traffic signal controller and the vehicle controller.
The overall system-level performance lies in the traffic and energy efficiency, safety, and mobility accessibility, such as all vehicles' travel time (delay) or average velocity, traffic congestion, and fuel consumption~\citep{balaji2010multi, wu2021flow}.
First, in the traffic signal control problem, an advantage of applying MAS is its (large-scale) problem decomposition capability stemming from its flexible and modular system structure in a hierarchical or fully distributed way~\citep{balaji2010multi}.
In particular, the concept of road network is employed, where an intersection is a node, the traffic signal timings of which are controlled by an (intersection) agent that, therefore, forms a MAS.
The conflict to be resolved in the coordination of multi-agent signal control can be not to further increase the congestion status of adjacent agents.

Second, an emerging direction in autonomous driving (AD) is to refocus on its multi-agent nature on top of ego-vehicle tasks like perception, recognition, localization, and maneuver control, such as works in connected autonomous vehicles.
One public benefit of investigating AD in the MAS perspective is to consider the transportation system as a whole, contribute to system-level traffic problems, and study their impacts on such as traffic flow, rather than the only standard of local or vehicle-level objectives~\citep{wu2021flow}.
Especially in the context of incompatible coordination protocols, the single-agent investigation perspective is not enough and autonomous driving techniques capable of coping with (heterogeneous) multi-agent interaction are inevitable, where agents are diverse, intelligent road users like pedestrians~\citep{rasouli2019autonomous}, human-driven vehicles, and autonomous vehicles of different automation levels from diverse manufacturers.
Examples of coordinated AD tasks are the vehicle platooning~\citep{li2015overview,liang2015heavy}, cruise control~\citep{desjardins2011cooperative}, merging~\citep{shalev2016safe}, navigating through traffic intersections~\citep{kalantari2016distributed}, adversarial traffic~\citep{wachi2019failure}, following, lane changing, and overtaking~\citep{yu2019distributed}.
For more coordinated task definition and approaches in AV, the readers can refer to the survey~\citep{mariani2021coordination}.

\subsection{Humanoid and Anthropomorphic Robot}
\label{sec_humanoid}




Humanoid robots and robots with anthropomorphic structures are a critical subfield of robotics.
They are needed to transfer human skills to robots for assisting or replacing people in such human-centered environments, hazardous scenes, space, or workplaces unreachable by humans~\citep{darvish2023teleoperation,ren2024enabling}.
Multi-agent coordination is vital in humanoid and anthropomorphic robots when a single agent is not capable of completing tasks, where an agent is a subsystem of the robot.
For example, multi-arm manipulators are more advantageous than a single manipulator in cases of, e.g., grasping relatively heavy, large, or long objects (more examples see~\citep{ollero2021past} and references therein).
Furthermore, multi-agent, especially decentralized, solutions are superior in scalability and adaptability performance, such as robustly handling unknown objects of arbitrary geometric and physical properties with malfunctioning components (agents) by various team sizes~\citep{ren2024enabling,tao2023multi,ha2021learning}.
The following examples of humanoid and anthropomorphic robots highlight the multi-agent coordination applied in a single robot.

For the dual-arm robot, a cooperative grasping means generating optimal grasp pairs based on the combination of grasp poses of each single-arm, i.e., the combinatorial optimization in the joint configuration space, whose dimensionality is determined by the number of independent degrees of freedom (DoF).
Successively, the spatio-temporal coordination in manipulation is achieved based on the motion synchronization that trajectories of multiple arms are synchronized in time at some physical positions for such as cooperatively lifting up objects or handover of items (i.e., the associated transition of object control)~\citep{asfour2006coordinated}.
Meanwhile, the intrinsic system safety in terms of inter-manipulator collision avoidance is prioritized in motion planning due to their highly overlapping workspaces, apart from the single-arm's capability to avoid collisions with environmental obstacles~\citep{das2020learning}.
Finally, the above coordination requirements can be integrated into one controller's design by, e.g., the multi-objective optimization~\citep{ren2024enabling}, or solved by multi-agent decentralized control for scalable multi-arm systems~\citep{ha2021learning}.

Besides, the dexterous (multi-fingered) (humanoid) robot hand is irreplaceable or advantageous than the (multi-)arm alone in manipulation tasks that need, such as compensation for the limited arm functionality or in-hand dexterity (e.g., fine manipulation other than grasping).
The definition of dexterity is often related to the generality concerning the set of tasks that can be accomplished and the overall system-level performance of manipulator(s)  (see discussions in~\citep{ma2011dexterity}).
For the anthropomorphic hand, dexterous tasks are typical to emulate coordinated human behaviors~\citep{bullock2011classifying,feix2015grasp}, which can be evaluated in a qualitative or open quantitative manner.
The dexterity is achieved by the kinematic redundancy brought by the multi-finger and their coordination.
The manipulation policy is usually centralized, i.e., one controller for all actuators~\citep{andrychowicz2020learning}, in spite of the many independently controlled/actuated DoF.
As with any centralized MAS, such centralized motion planners are vulnerable to individual component malfunction, are robot-structure-specific and object-specific~\citep{tao2023multi}, and suffer from poor scalability.
Accordingly, decentralized control of multi-agent reinforcement learning approaches emerge~\citep{wen2022multi,tao2023multi}, where a finger can be an agent.

Furthermore, a humanoid robot is a more complex multi-agent system that involves higher-order relationships of subsystems for a coordinated robot's reaction or decision-making.
For instance, a humanoid robot's coordination can refer to task decomposition and task transition between subsystems, such as in mobile manipulation tasks.
The coordinated head-eye movements of a humanoid robot can describe that the motions of the head and eye are compensatory in the active perception, e.g., the head in clockwise and the eye in anti-clockwise when looking at a fixed point in the scene~\citep{kuang2012active}.
Besides, the humanoid robot's coordination can happen between multi-modal sensors.
For example, the capability limitations of a moving robot's visual perception can be addressed by the acoustic (localization) system in tracking humans in the surroundings~\citep{asfour2006coordinated}.
The humanoid robotic tactile-sensing skin can be a network of multi-modal sensor modules with local controllers for preprocessing signals and routing data, which are then fused into robot reactions~\citep{mittendorfer2011humanoid}.
In this sense, it is a multi-agent system with sensor fusion coordination problems.





\subsection{Satellite Systems}
\label{sec_satellite}

Satellite services play a crucial role in the ongoing revolution of the space industry with the advent of New Space~\citep{nasanewspace2017,kodheli2020satellite}.
Distributed space systems (DSS) are naturally multi-agent systems (MAS), such as satellite formation flying (e.g., trailing), satellite cluster, satellite constellation, fractionated satellite, and satellite swarm~\citep{sundaramoorthy2010systematic,radhakrishnan2016survey}.
As agents in DSS, satellites can be classified based on the orbit's altitude and satellite mass, including Geostationary (GEO) / Medium Earth Orbit (MEO)  / Low Earth Orbit (LEO) / Very Low Earth Orbit (VLEO) satellites and small satellites: mini/micro/nano(cube)/pico/femto-satellite~\citep{radhakrishnan2016survey,kodheli2020satellite}.
The following brief surveys the MAS in three DSS.

First, in the constellation, satellites in different orbital planes coordinate to provide the desired service coverage, which may be global, regional, cellular, or demand-based.
The continuity of coverage service is optional due to the relative movement of non-GEO satellites to the Earth and the number of satellites.
The constellation design is typically a many-objective optimization or constrained optimization problem, which is evaluated by area coverage-based metrics (e.g., coverage probability), coverage gap-based metrics (e.g., revisit time), cost-based metrics (e.g, number of satellites, number of orbital planes or altitude), and communication-based metrics and constraints (e.g., latency, propagation path loss, radio resources, co-channel interference, noise level, and Doppler frequency offset/drift)~\citep{savitri2017satellite,singh2020low,al2021optimal}.
For instance, SpaceX proposes three technical modifications to its Starlink constellation configuration to facilitate its broadband internet service deployment together with increasing orbital debris mitigation and space safety (e.g., collision risk)~\citep{spacex2018,spacex2019,spacex2020}.

Second, the satellite swarm is an open, promising, and active research topic~\citep{radhakrishnan2016survey,kodheli2020satellite,farrag2021satellite}, which has attracted much researchers' attention since 2000s~\citep{engelen2011systems}.
As with other swarm systems, it is defined in terms of the key features of self-organization mechanism, coordinated behavior, and a common objective (distributed space mission)~\citep{sundaramoorthy2010systematic}.
This configuration is often related with the small satellite field due to the mass production and deployment requirements for a swarm, such as the constellation of 4 hierarchical swarms of total 28 small satellites in the Magnetic Nano-Probe Swarm mission~\citep{lubberstedt2005magnas}, the swarm of 36 cubesats in the QB50 project~\citep{QB50}, and the swarm of 50 nano-satellites in the radio telescope project OLFAR~\citep{engelen2010olfar}.
General multi-agent tasks and challenges apply in satellite swarm missions, such as the mission scheduling and planning~\citep{zheng2017swarm}, 
consensus~\citep{sarlette2007cooperative}, 
synchronization~\citep{marrero2022architectures}, 
remote sensing~\citep{farrag2021satellite}, 
collision avoidance~\citep{nag2013behaviour}, 
and target tracking and navigation (e.g., in asteroid characterization~\citep{stacey2018autonomous}).
However, the work in the satellite swarm is still limited, and more researches are expected to empower the swarm and demonstrate the swarm's advantageous properties in more scenarios.

Third, in satellite communications, many multi-agent (sequential) decision-making problems and coordination tasks are vital for domain innovation and advancement with the natural advantages of reconfigurable, scalable, upgradable, robust, and low complexity from distributed control.
For example, in optimizing the radio resources (spectrum, interference) and system performance (including load balancing), multi-agent beam hopping (scheduling) methods are proposed for the multi-beam satellite system~\citep{lin2022dynamic}, multi-satellite multi-beam system~\citep{zhu2021satellite,lin2024satellite}, and construction of phased array antenna~\citep{quadrelli2019distributed}, where an agent may control a beam parameter, a satellite's multiple beams, or a cubesat in a flying swarm.
Besides, routing in networks involving satellites is different from that in terrestrial networks due to the time-varying topology (including predictable regular changes and unpredictable link or node failures), frequent connection switching, and propagation delay arising from the satellites' moving and space environment.
An advantage of distributed routing protocols is their demonstrated scalability in thousands of satellites~\citep{zhang2021aser,lyu2024dynamic}.
Challenges in this direction include routing in hierarchical/vertical space networks with heterogeneous link connections and Space Internet with long distance and high delay communication links~\citep{kodheli2020satellite}.

\subsection{LLM-based Multi-Agent Systems}
\label{sec_mas_llm}

Large language model (LLM) agents have increasingly demonstrated human-level performance in a variety of activities, such as reasoning, planning, and problem-solving. Inspired by their surprising capabilities, various LLM-based multi-agent systems have been proposed to leverage the collective intelligence and specialized profiles and skills of different agents to solve more complex and challenging problems, e.g., software development, society simulation, and autonomous driving, in which multiple agents need to collaboratively engage in discussions, reasoning, and planning. This process mimics the intelligent behaviors of human/animal group work in problem-solving tasks. According to different application domains, existing work can be roughly categorized into two classes, including decision-making and behavior simulation. We elaborate on them below.

{\bf Decision-making.} The core idea behind using LLM-based MAS for decision-making is to harness the collective capabilities of multiple agents with diverse expertise. These agents, each acting as an expert, collaborate to address complex tasks efficiently, such as software development, embodied intelligence, and scientific research. \citet{li2023camel} introduces a novel communicative agent framework called CAMEL, which enables autonomous cooperation among agents through role-playing and inception prompting, aiming to achieve autonomously collaborative programming according to only users' text instructions. A concurrent work~\citep{qian2023communicative} shares the same spirit with~\citep{li2023camel}. Within the field of embodied intelligence, RoCo~\citep{mandi2024roco} develops a framework for multi-robot collaboration, which uses LLMs for both high-level communication and low-level motion planning, achieving better coordination in a variety of multi-robot tasks. ReAd~\citep{zhang2024towards} further improves coordination efficiency by using principled credit assignment of multiple agents. \citet{chen2023multi} develops an LLM-based consensus-seeking method that can be applied as a high-level planner for multi-robot coordination tasks. For scientific research, \citet{zheng2023chatgpt} automates chemical experiments using multiple LLM agents, each tasked with specific experimental and analytic tasks, including making plans, literature retrieval, experiment execution, and result summary. 

{\bf Behavior Simulation.} The second primary application scenario of LLM-based MAS is to simulate a variety of behaviors in different environments, e.g., social interaction~\citep{park2023generative}, game-playing, role-playing~\citep{ge2024scaling}, recommendation systems~\citep{agent4rec}, etc. The main advantages of leveraging LLM-based MAS for behavior simulation lie in their excellent in-context learning and instruction-following capabilities which are essential for vividly mimicking various behaviors and roles in different scenarios. Unlike decision-making problems that typically pay more attention to how cooperation happens among agents, behavior simulation often requires diverse methods for agent management, communication, and collaboration, because of the high complexity of the real-world scenarios and interactions. Next, we review behavior simulation conducted in various applications.

For social interaction, LLM-based MAS is adopted to simulate various behaviors and outcomes of the society, with the aim of exploring potential social dynamics and evolution, validating society experiments, and building virtual spaces and environments with realistic social phenomena. Early work by~\citep{park2023generative} develops generative agents in a simulation game environment, in which human users can engage with a modest community of multiple AI agents through natural language (NL) communication. On top of this groundbreaking work, \citet{gao2023s} construct large-scale social networks comprising 8,563 and 17,945 LLM agents, enabling complex and emerging social behavior simulation. Furthermore, recent research such as~\citep{ge2024scaling} also explores social interaction simulation to scale synthetic data generation for LLM training. 

LLM-based MAS is suitable for game-playing via different roles' behavior simulation, in which agents play diverse non-player characters (NPCs) within games and interact with human players. \citet{akata2023play} investigate the behavior of LLM agents in repeated social interactions using behavioral game theory and find the need for improved coordination and forgiveness strategies in LLMs to better align with human social conventions. \citet{mukobi2023welfare} introduces a new variant of the board game Diplomacy called Welfare Diplomacy to benchmark the cooperative capabilities of LLM agents. Built upon this, \citet{li2023theory} propose a multi-agent text game benchmarking the LLM agents' Theory of Mind. Moreover, there is a work~\citep{DBLP:conf/aaai/FanCJ024} that explores whether LLMs can act as rational players in game theory experiments. The authors systematically analyze the capabilities of LLM-based MAS in three key aspects of rationality: building clear desires, refining beliefs about uncertainty, and taking optimal action using three classical game environments including Dictator, Rock-Paper-Scissors, and Ring Network.
\section{Future and open topics}
\label{sec_challenge}

The multi-agent perspective offers the capabilities of modeling complex systems with complex interactions through problem decomposition, decentralized control and emerging swarm intelligence from distributed computational intelligence.
Despite substantial progress in the field of multi-agent systems (MAS) and coordination mechanisms, several key challenges remain, offering opportunities for future research.

\subsection{Scalability and Hybrid Coordination}
Scalability remains a critical challenge for future research in multi-agent coordination as the scale of the problem increases.
When the number of agents grows significantly, evaluating conflicts, clustering agents, and dynamically updating strategies all become increasingly complex.
Traditional methods may struggle to maintain efficiency and computational feasibility, leading to delays or suboptimal performance.
Therefore, how to efficiently tackle the large-scale coordination problem would be the first concern.
In turn, the multi-agent coordination is scalable if the (system-level) performance improves proportionately or will not degrade with the scale of agents, which motivates the large-scale deployment.
Then, how to coordinate and learn at scale by taking advantage of the scalable computational resources, information, or experiences (datasets) provided by a large amount of (distributed) agents would be another concern.

From the unified perspective, the hybridization of hierarchical and decentralized mechanisms is a practical solution and promising open direction.
First, hierarchical structures exist in nature and human behaviors. 
Multi-agent systems can incorporate hierarchical mechanisms as a crucial complementary or part of the self-organization. 
The hierarchical mechanism assigns greater responsibilities to some agents, allowing the MAS to adapt flexibly to the demands of the task at hand.
Therefore, hierarchy enables scalability and efficient management of large-scale MAS.
A similar idea is discussed by~\citet{mariani2021coordination} and~\citet{vasirani2011artificial} in autonomous driving that propose to select representative vehicles or software agents as regional leaders in the road (infrastructure) network taking charge of regional demands and altogether optimizing the overall traffic.
Besides, \citet{rizk2019cooperative} points out that hierarchical approaches promote the scalability by supporting diverse interaction densities from the local and global scales.
Second, decentralization enhances robustness and adaptability.
The hybrid approaches strike a balance between the strengths of hierarchical and decentralized approaches, providing the scalability and global coordination of a hierarchical system while maintaining the local adaptability and resilience of a decentralized one. 
This allows MAS to handle complex tasks with greater efficiency, ensuring swift local decision-making while maintaining overall system coherence, making it ideal for large and dynamic environments.

\subsection{Heterogeneity and Human-MAS Coordination}
Most existing MAS are homogeneous and have a limited ability to solve complex tasks. 
In addition, to solve tasks more efficiently, specialization is very essential. 
Thus, we believe that heterogeneous MAS (including physical, behavioral, or logical heterogeneity) with specialized agents is a promising research direction. 
In such a heterogeneous MAS, agents play different roles and have different capabilities.
This brings an additional degree of freedom for the flexibility in coordination.
Recently, substantial experimental results have been achieved by LLM-based heterogeneous MAS in solving various language-based AI tasks~\citep{zhuge2023mindstorms, li2023camel, qian2023communicative, park2023generative, gao2023s, ge2024scaling}.
However, heterogeneity means that not all agents coordinate in the same way as in homogeneous MAS.
This challenges the coordination of heterogeneous MAS, since homogeneity may be the simplest mechanism for forming self-organized patterns.
In particular, humans, as a kind of agent with incentives and emotions, play an important role and can be passively or proactively involved in coordination, which can be respectively called the human-MAS interaction and human-MAS teaming in this context.

For the human-MAS interaction, for example, mixed traffic is a passive coordinated heterogeneous MAS, where pedestrians, bicycle riders, motorcycle riders, horse riders, human-driven vehicles, partial and fully autonomous vehicles are passively accounted for by each other since they are heterogeneous with respect to their coordination protocols, yet share the road with some common goals like keeping mutual mobility and safety.
The research on autonomous driving (AD) has investigated being aware of and paying attention to human road users through scene representation~\citep{chitta2021neat} and investigated the impacts of applying AD.
But how to effectively coordinate with heterogeneous agents regarding complex inter-agent dependencies has not been extensively studied in coordinated multi-agent AD, especially under the system-level considerations that will not break societal norms and expectations by introducing  autonomous vehicles, apart from the whole system efficiency, fairness, emergency priority, etc. 

For the human-MAS teaming, human agents proactively engage in coordinated decision-making with AI agents by teleoperation or supervisory control.
It is typically termed the human-swarm interaction (HSI) in the literature~\citep{kolling2015human,dorigo2021swarm}.
This human-centric heterogeneous MAS fuses the abstract human intelligence and enables the injection of human intent into AI MAS, extends the potential of each other and exploits the complementary capabilities of humans and machines.
Such solutions are crucial for the fifth industrial revolution and future MAS advantages.
Many human-MAS interaction techniques have been explored (see~\citep{kolling2015human, vaidis2021swarm, lin2022modelling, lin2023adaptive, aldini2022detection, wang2022implicit, john2024prediction}), including graphical interfaces, gesture recognition, touch feedback, voice recognition, eye movement, etc. 
Particularly, the natural brain-computer interface (nBCI)~\citep{aldini2022detection, john2024prediction, wang2022implicit} has emerged as a promising technology for natural and hands-free interaction, especially in applications involving wearable devices. 
nBCI leverages neural signals to enable seamless communication between humans and machines, thereby enhancing interaction efficiency in human-MAS systems. Several studies have demonstrated the potential of nBCI in human-MAS interaction. 
For example, ~\citet{wang2022implicit} proposed a novel error-related potential (ErrP)-based nBCI for implicit robot control, where human responses to unexpected robot actions were translated into corrective commands, enabling real-time closed-loop interaction without explicit input. 
Additionally, ~\citet{aldini2022detection, john2024prediction} integrated cognitive conflict detection based on EEG signals into physical human-robot collaboration to assess human responses under unexpected robot behaviors, demonstrating that nBCI can effectively monitor human cognitive states to support adaptive robot control. 
These studies collectively highlight the potential of nBCI to enhance human-MAS interaction by providing real-time cognitive feedback for adaptive decision-making in dynamic environments.

Last, the key to maximize the above potential is the superior coordination process with two promising and challenging directions in co-learning and trustworthy. 
Computational trust modeling \citep{lin2023adaptive, lin2022modelling, wang2023robot} has been proposed to facilitate reliable and adaptive collaboration between human agents and machine agents in human-MAS systems. Trust modeling involves assessing the trustworthiness of agents and adapting interactions accordingly to ensure efficient and safe collaboration. 
For example,~\citet{lin2022modelling, lin2023adaptive} introduced trust models based on reinforcement-learning-based fusion, enabling MAS to dynamically adjust their behaviours according to human feedback and past experiences. 
~\citet{wang2023robot} investigated trust-based role arbitration in human-robot teams, where trust levels determine role assignments and decision-making authority, thereby enhancing collaboration under dynamic conditions. These advancements in computational trust modeling are pivotal for human-centric MAS, as they not only improve team performance but also foster human acceptance and confidence in MAS technologies.

\subsection{Learning and LLM-based MAS}
Recent LLM-based MAS such as GPTSwarm~\citep{zhugegptswarm}, MetaGPT~\citep{hong2024metagpt} and swarm\footnote{\url{https://github.com/openai/swarm}} have shown the ability to solve different tasks. Popular LLM-based MAS frameworks include AutoGen\footnote{\url{https://microsoft.github.io/autogen/0.2/}}, LangChain\footnote{\url{https://github.com/langchain-ai/langchain}}, AutoGPT\footnote{\url{https://github.com/Significant-Gravitas/AutoGPT}}, LangGraph\footnote{\url{https://langchain-ai.github.io/langgraph/}}, CrewAI\footnote{\url{https://www.crewai.com/}}, etc. Looking ahead, building applications upon LLM-based MAS will play a crucial role in both academia and industry. Furthermore, the LLM-based MAS will make AI agents more intelligent and helpful in human society. We believe that the LLM-based MAS is a trend. This is only the beginning, with much more exciting progress to come.

In essence, the working functionality of LLMs is to fit the complex statistical model into the huge training dataset or joint state space for MAS. 
As a result, like all other machine learning techniques, LLMs can also suffer from the poor generalization ability (e.g., hallucination) if the training dataset or joint state space cannot cover the cases of interest well. 
Furthermore, the training process of LLMs is often expensive in terms of both economic and labor costs.

\section{Conclusion}\label{sec_conclusion}


The MAS perspective and coordination capability have already reformed and will further revolutionize diverse applications.
This survey provides a unified insight into the interdisciplinary MAS coordination by analyzing and answering four fundamental questions: (1) what is coordination; (2) why coordination; (3) who to coordinate with; and (4) how to coordinate.
Particularly, three general MAS coordination problems and six MAS applications are surveyed, which cover fundamental common tasks, widely studied domains, and newly emerged areas.
Given on the above efforts, three future directions: hybridization of hierarchical and decentralized coordination, human-MAS coordination, and LLM-based MAS are explored for three open MAS performance: scalability, heterogeneity, and learning mechanism.
We anticipate that multi-agent coordination will drive a new stage of general AI.

\section{Acknowledgements}

This work is partially supported by 
the National Natural Science Foundation of China under Grant No. 72401237 and No. 61761136008,
the Shenzhen Fundamental Research Program under Grant No. JCYJ20200109141235597,
and the Australian Research Council (ARC) under Discovery Grant No. DP180100656 and No. DP210101093.

\bibliographystyle{ACM-Reference-Format}
\bibliography{mybibfile}

\end{document}